\documentclass[useAMS,usenatbib]{mn2e}
\usepackage{graphicx}
\title[SPIRE detection of high-z massive compact galaxies in GOODS-N field]{\vspace{-0.8cm} HerMES : SPIRE detection of high redshift massive compact galaxies in GOODS-N field}
\author[A.~Cava et al.]
{\parbox{\textwidth}{\vspace{-0.5cm} \Large A.~Cava,$^{1,2}$\thanks{E-mail: \texttt{acava@iac.es}}
G.~Rodighiero,$^{3}$
I.~P{\'e}rez-Fournon,$^{1,2}$
F.~Buitrago,$^{4}$
I.~Trujillo,$^{1,2}$
B.~Altieri,$^{5}$
A.~Amblard,$^{6}$
R.~Auld,$^{7}$
J.~Bock,$^{8,9}$
D.~Brisbin,$^{10}$
D.~Burgarella,$^{11}$
N.~Castro-Rodr{\'\i}guez,$^{1,2}$
P.~Chanial,$^{12}$
M.~Cirasuolo,$^{13}$
D.L.~Clements,$^{12}$
C.J.~Conselice,$^{4}$
A.~Cooray,$^{6,8}$
S.~Eales,$^{7}$
D.~Elbaz,$^{14}$
P.~Ferrero,$^{1,2}$
A.~Franceschini,$^{3}$
J.~Glenn,$^{15}$
E.A.~Gonz\'alez~Solares,$^{16}$
M.~Griffin,$^{7}$
E.~Ibar,$^{13}$
R.J.~Ivison,$^{13,17}$
L.~Marchetti,$^{3}$
G.E.~Morrison,$^{18,19}$
A.M.J.~Mortier,$^{12}$
S.J.~Oliver,$^{20}$
M.J.~Page,$^{21}$
A.~Papageorgiou,$^{7}$
C.P.~Pearson,$^{22,23}$
M.~Pohlen,$^{7}$
J.I.~Rawlings,$^{21}$
G.~Raymond,$^{7}$
D.~Rigopoulou,$^{22,24}$
I.G.~Roseboom,$^{20}$
M.~Rowan-Robinson,$^{12}$
D.~Scott,$^{25}$
N.~Seymour,$^{21}$
A.J.~Smith,$^{20}$
M.~Symeonidis,$^{21}$
K.E.~Tugwell,$^{21}$
M.~Vaccari,$^{3}$
I.~Valtchanov,$^{5}$
J.D.~Vieira,$^{8}$
L.~Vigroux,$^{26}$
L.~Wang$^{20}$ and
G.~Wright$^{13}$}\vspace{0.4cm}\\
\parbox{\textwidth}{\raggedright $^{1}$Instituto de Astrof{\'\i}sica de Canarias (IAC), E-38200 La Laguna, Tenerife, Spain\\
$^{2}$Departamento de Astrof{\'\i}sica, Universidad de La Laguna (ULL), E-38205 La Laguna, Tenerife, Spain\\
$^{3}$Dipartimento di Astronomia, Universit\`{a} di Padova, vicolo Osservatorio, 3, 35122 Padova, Italy\\
$^{4}$School of Physics and Astronomy, University of Nottingham, NG7 2RD, UK\\
$^{5}$Herschel Science Centre, European Space Astronomy Centre, Villanueva de la Ca\~nada, 28691 Madrid, Spain\\
$^{6}$Dept. of Physics \& Astronomy, University of California, Irvine, CA 92697, USA\\
$^{7}$Cardiff School of Physics and Astronomy, Cardiff University, Queens Buildings, The Parade, Cardiff CF24 3AA, UK\\
$^{8}$California Institute of Technology, 1200 E. California Blvd., Pasadena, CA 91125, USA\\
$^{9}$Jet Propulsion Laboratory, 4800 Oak Grove Drive, Pasadena, CA 91109, USA\\
$^{10}$Space Science Building, Cornell University, Ithaca, NY, 14853-6801, USA\\
$^{11}$Laboratoire d'Astrophysique de Marseille, OAMP, Universit\'e Aix-marseille, CNRS, 13388 Marseille cedex 13, France\\
$^{12}$Astrophysics Group, Imperial College London, Blackett Laboratory, Prince Consort Road, London SW7 2AZ, UK\\
$^{13}$UK Astronomy Technology Centre, Royal Observatory, Blackford Hill, Edinburgh EH9 3HJ, UK\\
$^{14}$Laboratoire AIM-Paris-Saclay, CEA/DSM/Irfu - CNRS - Universit\'e Paris Diderot, CE-Saclay, F-91191 Gif-sur-Yvette, France\\
$^{15}$Dept. of Astrophysical and Planetary Sciences, CASA 389-UCB, University of Colorado, Boulder, CO 80309, USA\\
$^{16}$Institute of Astronomy, University of Cambridge, Madingley Road, Cambridge CB3 0HA, UK\\
$^{17}$Institute for Astronomy, University of Edinburgh, Royal Observatory, Blackford Hill, Edinburgh EH9 3HJ, UK\\
$^{18}$Institute for Astronomy, University of Hawaii, Honolulu, HI 96822, USA\\
$^{19}$Canada-France-Hawaii Telescope, Kamuela, HI, 96743, USA\\
$^{20}$Astronomy Centre, Dept. of Physics \& Astronomy, University of Sussex, Brighton BN1 9QH, UK\\
$^{21}$Mullard Space Science Laboratory, University College London, Holmbury St. Mary, Dorking, Surrey RH5 6NT, UK\\
$^{22}$Space Science \& Technology Department, Rutherford Appleton Laboratory, Chilton, Didcot, Oxfordshire OX11 0QX, UK\\
$^{23}$Institute for Space Imaging Science, University of Lethbridge, Lethbridge, Alberta, T1K 3M4, Canada\\
$^{24}$Astrophysics, Oxford University, Keble Road, Oxford OX1 3RH, UK\\
$^{25}$Department of Physics \& Astronomy, University of British Columbia, 6224 Agricultural Road, Vancouver, BC V6T~1Z1, Canada\\
$^{26}$Institut d'Astrophysique de Paris, UMR 7095, CNRS, UPMC Univ. Paris 06, 98bis boulevard Arago, F-75014 Paris, France}}

\begin{document}
\label{firstpage}
\date{Accepted ..... Received .....; in original form .....}
\pagerange{\pageref{firstpage}--\pageref{lastpage}} \pubyear{2010}
\maketitle
\begin{abstract} {\small
We have analysed the rest-frame far infrared (FIR) properties of a sample of massive ($M_{\star}>10^{11}$M$_{\odot}$) galaxies at $2\la z \la3$ in the GOODS (Great Observatories Origins Deep Survey) North field using the Spectral and Photometric Imaging Receiver (SPIRE, Griffin et al. 2010) instrument aboard the {\it Herschel Space Observatory}. To conduct this analysis we take advantage of the  data from the HerMES key program. The sample comprises 45 massive galaxies with structural parameters characterised  with {\it HST} NICMOS-3.  We study detections at submm {\it Herschel} bands, together with {\it Spitzer} 24$\mu$m data, as a function of the morphological type, mass and size. We find that 26/45 sources are detected at MIPS-24$\mu$m and 15/45 (all MIPS-24$\mu$m detections) are detected at SPIRE-250$\mu$m, with disk-like galaxies more easily detected.
We derive star formation rates (SFR) and specific  star formation rates (sSFR) by fitting the spectral energy distribution (SED) of our sources, taking into account non-detections for SPIRE and systematic effects for MIPS derived quantities. We find that the mean SFR for the spheroidal galaxies ($\sim 50-100$M$_{\odot}$yr$^{-1}$) is substantially (a factor $\sim 3$) lower than the  mean value presented by disk-like galaxies ($\sim 250-300$M$_{\odot}$yr$^{-1}$).\\}
\end{abstract}
\section{Introduction}
One of the most intriguing recent results in extragalactic astrophysics is the discovery that massive galaxies ($M> 10^{11}$M$_\odot $) at high redshift were {\bf on average} more compact than their local counterparts (Daddi et al. 2005; Trujillo et al. 2006, 2007; Longhetti et al. 2007, Buitrago et al. 2008, B08 hereafter).  Although the formation mechanism for such compact galaxies is not yet perfectly established (e.g. Khochfar \& Silk 2006, Ricciardelli et al. 2010), the proposed scenario that drives these compact galaxies into the local population of massive ellipticals likely includes dry major mergers or, the currently favoured scenario, minor mergers with less dense galaxies (Naab et al. 2007; Hopkins et al. 2009).

In order to shed light on how these compact galaxies have formed the bulk of their mass, the characterisation of their star formation rate (SFR) history is required. While their morphological properties seem to be quite simple, their intrinsic physical properties are still not well characterised. Different groups have found contrasting results on the correlation between the SFR and structural properties. Kriek et al. (2009), for example, claim a striking correspondence between the SFR and structural parameters (sizes, morphologies), with the compact red galaxies forming stars at a rate one order of magnitude lower than larger blue galaxies. 
In contrast, P\'erez-Gonz\'alez et al. (2008, PG08 hereafter) using extremely deep 24~$\mu$m imaging of the Extended Groth Strip found that at z~$\sim2$, $\sim80\%$ of the disk-like galaxies and $\sim50\%$ of the compact spheroid-like galaxies are detected, with similar mean SFRs of $60-100$M$_{\odot}$yr$^{-1}$. However, within each morphological category the size is not related to the 24~$\mu$m detection.

In the present work we focus on a sample of massive ($M _{\star}\simeq 10^{11}-10^{12}$M$_\odot $) high redshift ($z\simeq 2-3$)  galaxies having  deep {\it HST} data with sizes and structural properties measured by B08 and examine their FIR properties, as measured from {\it Herschel {\footnote{{\it Herschel} is an ESA space observatory with science instruments provided by European-led Principal Investigator consortia and with important participation from NASA.}}}(Pilbratt et al. 2010) within the Herschel Multi-tiered Extragalactic Survey (HerMES, Oliver et al. 2010b),  as a function of morphology and compactness. This paper represents the first exploratory step in this direction.

Throughout this paper we assume a cosmology with H$_0=70\,$km$\,$s$^{-1}\,$Mpc$^{-1}$, 
$\Omega_{\rm M} = 0.3$, $\Omega_{\Lambda} = 0.7$. 
\section[]{The data}
The parent sample of galaxies originates from the work of B08. It covers the GOODS North and South fields and has been imaged as part of the  GOODS NICMOS Survey (GNS; PI C. Conselice). We make use only of the Northern sub-sample in this work.  

We take advantage of the HerMES science demonstration phase (SDP) observations carried out at 250, 350 and 500 $\mu$m, which cover a large ($\simeq$ 30~arcmin by 30~arcmin) area centred in the GOODS-N field. The instrument characteristics and capabilities are described in Griffin et al. (2010).  

\begin{figure*}
\includegraphics[width=1\columnwidth]{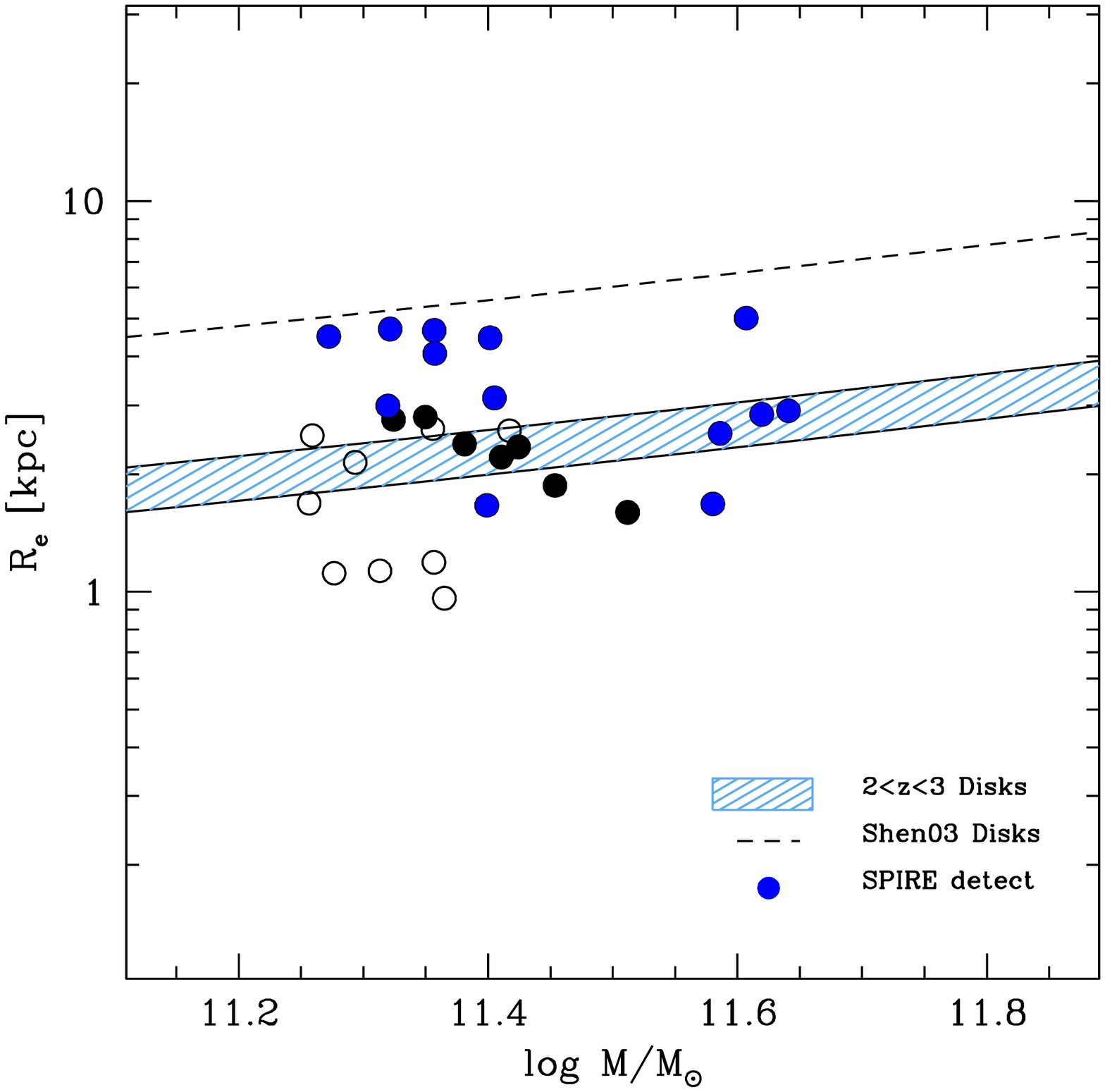}
\includegraphics[width=1\columnwidth]{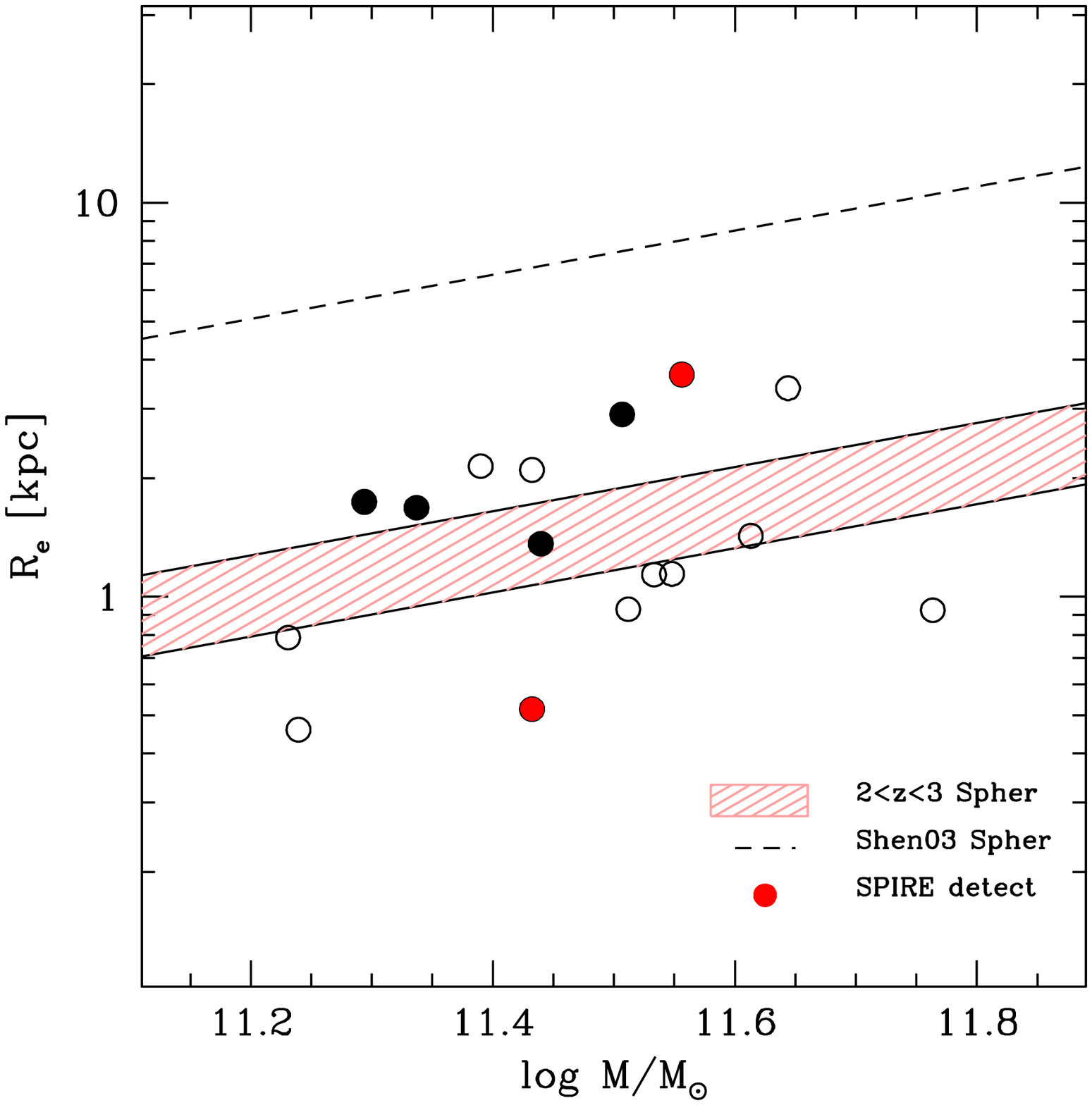}
\caption{Stellar mass-size relation for disk-like (left panel) and spheroid-like (right panel) galaxies. The dashed line in both panels indicates the local  relation (Shen et al. 2003). The shaded area represents the stellar mass-size relation for galaxies between $z=2$ (upper envelope) and $z=3$ (lower envelope) from B08. Black circles represents galaxies in the parent catalogue, all filled circles are 24$\mu$m detections, while coloured filled-circles (blue and red for disk-like and spheroid-like galaxies respectively) indicate the 250$\mu$m detection.} 
\label{MR_rel}
\end{figure*}

\subsection{The GNS sample}
The GNS is a large {\it HST} NICMOS-3 camera program of 60 pointings  centred around massive galaxies at $z \sim 1.7-3$ in the F160W ({\it H}) band in the GOODS North and South fields.  
The pointings were optimized to observe as many high-mass M$_{*}$ $> 10^{11}$ M$_{\odot}$ galaxies as possible, with the selection  of these targets described in Conselice et al. (2010).  

Stellar masses  and photometric redshift estimates for this sample were derived exploiting  multi wavelength ({\it BVRIizJHK}) data from GOODS (Giavalisco et al. 2004), using  Bruzual \& Charlot (2003) stellar population synthesis models, assuming a Chabrier (2003) Initial Mass Function (IMF). Only four spectroscopic redshifts are available for GOODS-N (Barger et al. 2008), showing a mean $\Delta z/(1+z)\sim 0.08$ with respect to the photometric redshifts. Details of  SED fitting techniques are in Conselice et al. (2007). 
As explained in the following section, in deriving the SFR we are implicitly assuming a Salpeter (1955) IMF. Therefore, for consistency and in order to derive the specific star formation rates (sSFR), we report all the masses used in this work to a Salpeter IMF. To scale the stellar masses of the parent catalogue from Chabrier to Salpeter we used the following equation: 
log$M_{\star}$(Salpeter)=log$M_{\star}$(Chabrier)$ + 0.23$
 and, when needed (for the Shen et al. 2003 stellar mass-size relation), we also scale from the Kroupa (2001) IMF to Salpeter using:
log$M_{\star}$(Salpeter)=log$M_{\star}$(Kroupa)$ +0.19$
(see e.g. Cimatti et al. 2008).
 Structural parameters and sizes  for the GNS sample were measured by using the \textsc{ GALFIT} code. The light distribution has been modelled with a S\'ersic model, deriving  S\'ersic indicx $n$, axis ratio $b/a$ and effective radius along the semi-major axis $a_{\epsilon}$. Our measured sizes are circularized so that $r_{0}=a_{\epsilon}\sqrt{1-\epsilon}$ with $\epsilon$ the projected ellipticity of the galaxy.  Details on the fitting procedure and reliability of the measurements can be found in B08 and Trujillo et al. (2007).

This data set, containing 45 galaxies for GOODS-N, represents the largest sample of high-$z$ massive galaxies to date, that have high-quality {\it HST} (NICMOS) data and measured structural parameters. This allows us to look for the first time at the possible correlations between rest-frame far IR properties, exploiting the available {\it Herschel} data, and the structural parameters (sizes and S\'ersic indices) of these high-$z$ massive galaxies. We will refer to this catalogue as the {\it parent catalogue} throughout the paper.
\subsection{The {\it Herschel} data}
The SDP GOODS-N observations are among the deepest possible with SPIRE, since the instrumental noise is lower then the confusion noise from overlapping background sources.  In building the catalogues two different approaches have been used. In one approach blind source extraction was performed, resulting in single-band catalogues at 250, 350 and 500$\mu$m (Smith et al. 2010, in prep.). In the second approach, flux densities and S/N ratios were obtained from PSF fitting on 24$\mu$m priors using a new source extraction method developed for HerMES called 'XID' (for details see Roseboom et al. 2010). 
Cross identifications are  performed in map space so as to minimize source blending effects and allow the recovery of the faintest sources, close to the confusion limit. 

Within a field of approximately 10~arcmin by 15~arcmin, which has been observed with {\it Spitzer} and {\it HST}, there are a total of $\sim$1500 24 $\mu$m sources which meet the reliability criterion of S/N$\ge$5 and $S_{24}\ge20 \mu$Jy.  {\it Herschel} detects about a third of them in at least one band (as reported by Elbaz et al. 2010).  

Following the approach of Elbaz et al. (2010), we use SPIRE measurements down to the $5\sigma$ limits of the prior catalogue of 4.4 mJy at 250$\mu$m. We note that these measurements lie close to the SPIRE confusion noise of 5.8 mJy (Nguyen et al. 2010).  However, this limit is a spatially averaged statistical limit which considers that galaxies are homogeneously distributed in the field and all affected in the same way by close neighbours. We take advantage of the higher resolution 24 $\mu$m image to flag 'clean' galaxies (for which SPIRE flux densities can be potentially more robust) by requiring that that there is at most one bright neighbour within 20 arcsec (close to the full width half maximum, FWHM, of the SPIRE 250 $\mu$m band) with 24 $\mu$m flux higher than 50 per cent of the central 24 $\mu$m counterpart.
By matching the resulting catalogue with the GNS catalogue from B08 we obtain a sample of 15 detected sources in total. Additionally we have 11 24$\mu$m priors without a 'clean' SPIRE detection. Of the 15 detections, 10 galaxies show detection at 350~$\mu$m and only 3 have detection at  500~$\mu$m.  Due to the large PSF of SPIRE and the associated deblending problems, especially at longer 350 and 500~$\mu$m  wavelengths, we focus our attention in this work on the 250~$\mu$m detections.  
As an additional check we also looked for possible 250$\mu$m detections not associated with the prior 24$\mu$m sources. 
We can confirm that none of the 19 GNS galaxies without 24~$\mu$m detections have a clear detection at 250~$\mu$m. 

\section{Analysis and results}
\subsection{IR luminosities and SFR}
We have performed an SED fitting procedure in order to derive the total IR luminosity of each source, by comparing the optical-to-IR photometry with a library of template SEDs of local objects from Polletta et al. (2007) and adding a few modified templates  (Gruppioni et al. 2010).  We have separately analyzed objects with SPIRE and/or MIPS detections. When limited to the 24 $\mu$m band, large extrapolations are required. However, with our technique (see Rodighiero et al. 2010) the inclusion of the whole SED in the fitting procedure allows us to fully exploit the photometric information and to determine appropriate k-corrections.   As described in Rodighiero et al. (2010), the advent of {\textit Herschel} data constrains the far-IR normalization of the SED, providing more accurate bolometric luminosities (and consequently SFR values). In both cases,  we forced the spectral fit to reproduce the far-IR  {\it Spitzer} and {\it Herschel} data points by weighting more the FIR points. 

As a measure of the total IR luminosity, our adopted procedure has been to integrate for each object the best-fit SED, in the rest-frame [8-1000$\mu$m] interval. For the galaxies detected with SPIRE or at 24~$\mu$m, the instantaneous SFR was then estimated using the calibration of Kennicutt et al. (1998):
SFR[$M_{\odot}$yr$^{-1}] = 1.7 \times 10^{-10}L([8 - 1000 \mu$m$])/$L$_{\odot}$, under the assumption that the bulk of the bolometric emission is dominated by star-formation with a Salpeter IMF. We have also checked that the galaxies in our sample do not have X-ray counterparts, ensuring the absence of optically thin AGNs detected by the {\it Chandra Space Observatory}. 
\begin{table*}
\centering
\caption{SPIRE 250~$\mu$m and MIPS 24~$\mu$m detection fraction, mean SFR and sSFR.}
\begin{tabular}{cccccccc}
\hline \hline
&  &\multicolumn{3}{|c|}{SPIRE 250~$\mu$m }&\multicolumn{3}{|c|}{MIPS 24~$\mu$m } \\
\hline 
 & N$_{{\rm tot}}$ & Detection & SFR[M$_{\odot}$yr$^{-1}$] & sSFR[Gyr$^{-1}$] & Detection & SFR[M$_{\odot}$yr$^{-1}$]& sSFR[Gyr$^{-1}$] \\
\hline 
All			& 45 & 33 $\pm$  8 \%	& 240 $\pm$ 50	& 1.05 $\pm$ 0.33	& 58 $\pm$ 11 \%	& 225$\pm$60& 0.95 $\pm$ 0.25	\\ 
n$> 2$		& 16 & 13 $\pm$  9 \%	& 95 $\pm$ 35	& 0.34 $\pm$ 0.13	& 38 $\pm$ 15 \%	& 45  $\pm$25& 0.17 $\pm$ 0.11	\\
n$< 2$		& 29 &  45 $\pm$ 12 \%	& 310 $\pm$ 45	& 1.25 $\pm$ 0.25	& 69 $\pm$ 15\%	& 280$\pm$70& 1.22 $\pm$ 0.30	\\
\hline \hline
\end{tabular}
\label{tab1}
\end{table*}
\subsection{Results}
We split our sample into early- (or spheroid) and late-type (or disk-like) galaxies, according to their measured S\'ersic indices in {\it H} band from the NICMOS images. B08 showed that a value of $n=2$  is optimal  to separate galaxies at these high redshifts into the two morphological classes when using NICMOS data. Therefore, we define spheroids as galaxies with $n>2$ and disk-like galaxies those with $n \le 2$. We obtain for the parent catalogue 29 disk-like galaxies and 16 spheroids (that is $\sim65$ per cent of the sample is composed of disk-like galaxies, while $\sim35$ per cent are spheroids). We have also checked, through one and two dimensional Kolmogorov-Smirnov tests, that the underlying distribution of stellar masses and redshift are compatible with being extracted from the same parent distribution (at a 99 per cent of confidence level). This should insure the detection fractions are not largely biased by the underlying distributions.

As an additional criterion, we split our sample according to the surface mass density (defined as $D = 0.5\cdot M/(\pi r_{0}^2)$) of our galaxies, adopting as threshold the value of $D_{{\rm t}} = 9\times 10^{9}$M$_{\odot}$kpc$^{-2}$, that is the median value for our sample. The resulting fraction of MIPS detected over-dense (with respect to the median value) galaxies result to be $\sim41\pm13\%$, while for under-dense galaxies is $\sim74\pm18\%$. Analogously, the SPIRE detected fractions result to be $\sim18\pm9\%$ and $\sim48\pm14\%$ for over-/under-dense galaxies respectively. In both cases, relatively less dense galaxies result to have higher detection fractions.

We have also checked through the Fisher one-sided exact test that the detected fractions are in agreement with the hypothesis that there is a connection between detection and morphology (at a $\sim98$ per cent confidence level).

As a first result in Fig.\ref{MR_rel}, we show the stellar mass-size relation for disk-like (left panel) and spheroid (right panel) galaxies, including MIPS and SPIRE detections. The dashed lines in each panel indicate the local mass-size relation (Shen et al. 2003). The shaded area represents the region enclosed between the  mass-size relation for high redshift galaxies at $2\la z \la 3$ following the evolution found in B08. All circles represent galaxies in the parent  catalogue, with filled circles (black and coloured) being 24$\mu$m detections and those in colour (blue and red for disk-like and spheroid-like galaxies respectively) being the 250$\mu$m detections. Both relations, local and high redshift, have been transformed to the Salpeter IMF using the equations reported in Section~2.1.

We note that SPIRE detections correspond mainly to larger and more massive disk-like galaxies ($\sim45$ per cent of the sample has a 250$\mu$m counterpart, and $\sim70$ per cent has a MIPS detection, see Fig.\ref{MR_rel}), while spheroids are mostly undetected (only $\sim13$ per cent have 250$\mu$m detection and only $\sim38$ per cent are MIPS detected). 
The mean radius for 250$\mu$m-detected disk-like galaxies in our sample is log($r_{0}$)=$0.50\pm0.10$, while for non-detected sources it is log($r_{0}$)=$0.37\pm0.08$. Furthermore, the mean mass for 250$\mu$m-detected disk-like galaxies in our sample is log(Mass)=$11.45\pm0.10$, while for non-detected sources it is log(Mass)=$11.40\pm0.05$. This results reflect in the trends shown in Fig.2 and are in agreement with previous sub-mm studies where it is found that the radius of the $850\mu$m-detected sources is larger than average optically-detected galaxies at essentially all redshifts (e.g. Chapman et al. 2003, Pope et al. 2005). In the case of spheroid-like galaxies a higher detection fraction is also suggested for larger objects, but this is not so clear, due to the small statistics.
The two spheroids detected with SPIRE have a Sersic index just above the threshold value (2.2 and 2.35). These are classified using a fit to the NICMOS imaging, but from inspecting the ACS images there is at least some hint of possible structure associated with one of the two spheroids. Additionally, we note that all the galaxies with a mass-size relation in better agreement with the local one (see also Fig.\ref{MR_rel}; left panel) could correspond to interacting systems, as suggested by the inspection of ACS images. If confirmed, large sizes would not indicate a big galaxy, but rather a compact galaxy in a merging phase. It is beyond the goals of the present Letter to examine the details of these objects, that are instead deferred to a future work.

\begin{figure}
\includegraphics[width=0.48\columnwidth]{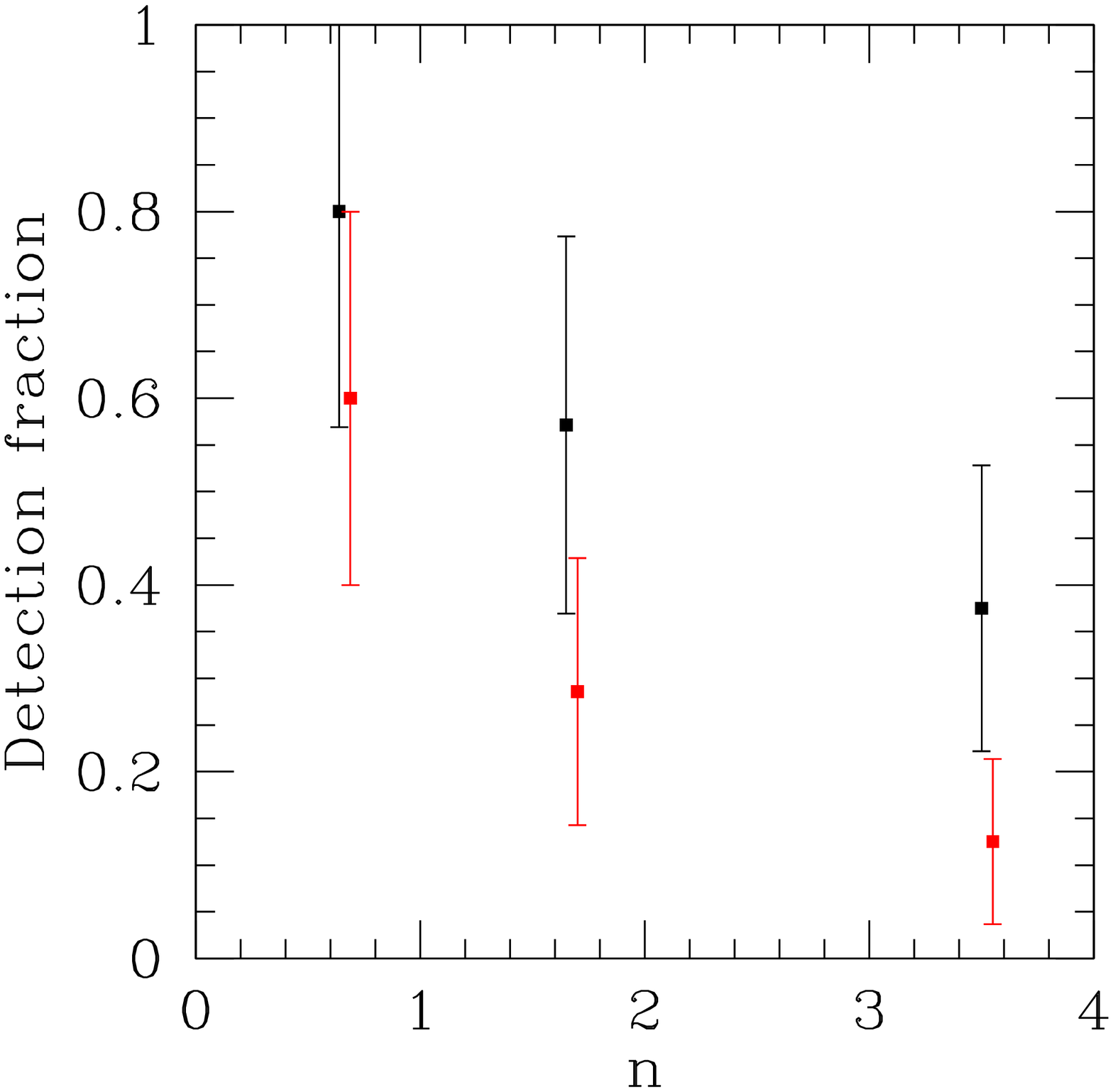}
\includegraphics[width=0.48\columnwidth]{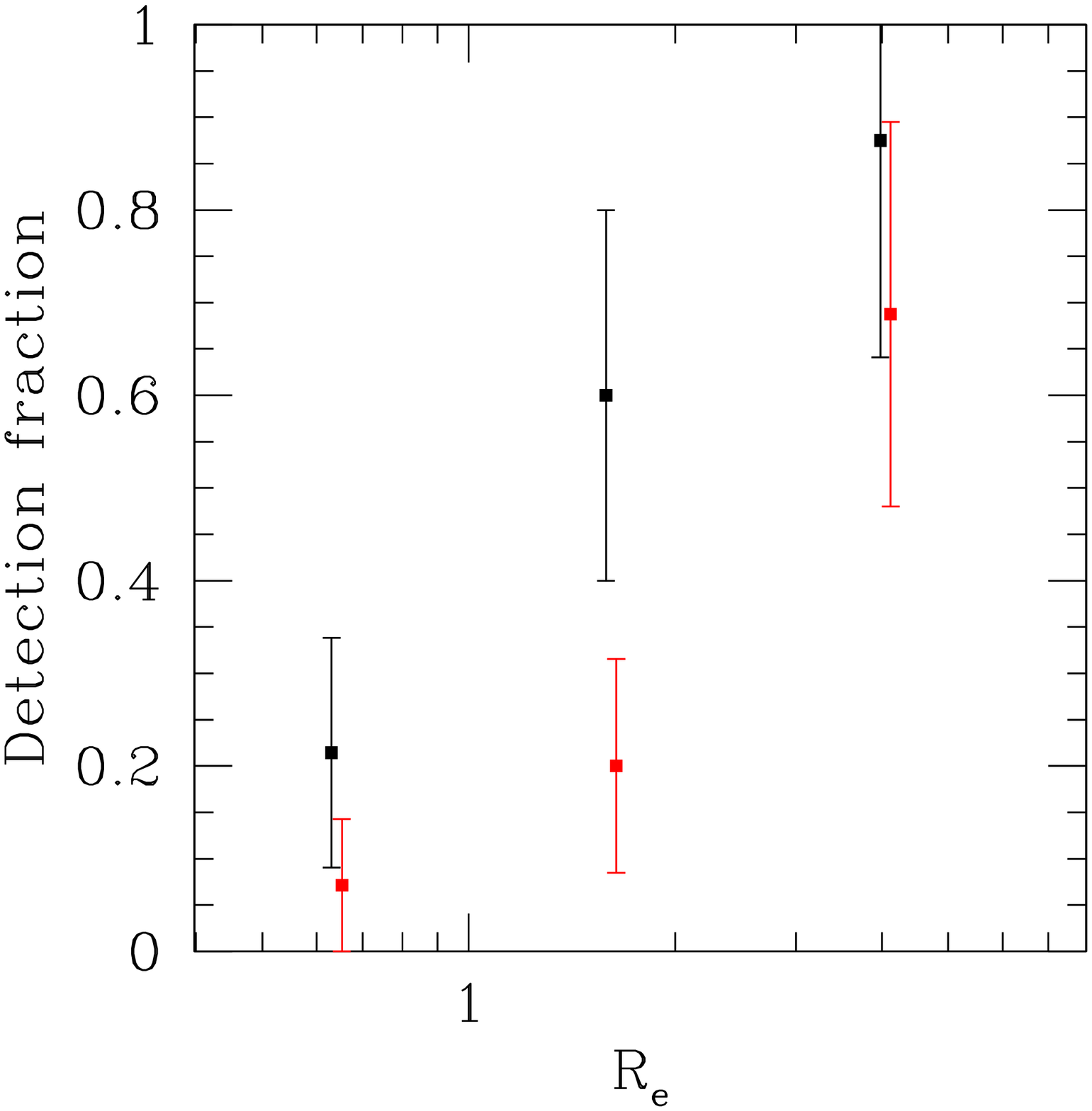}\\
\includegraphics[width=0.48\columnwidth]{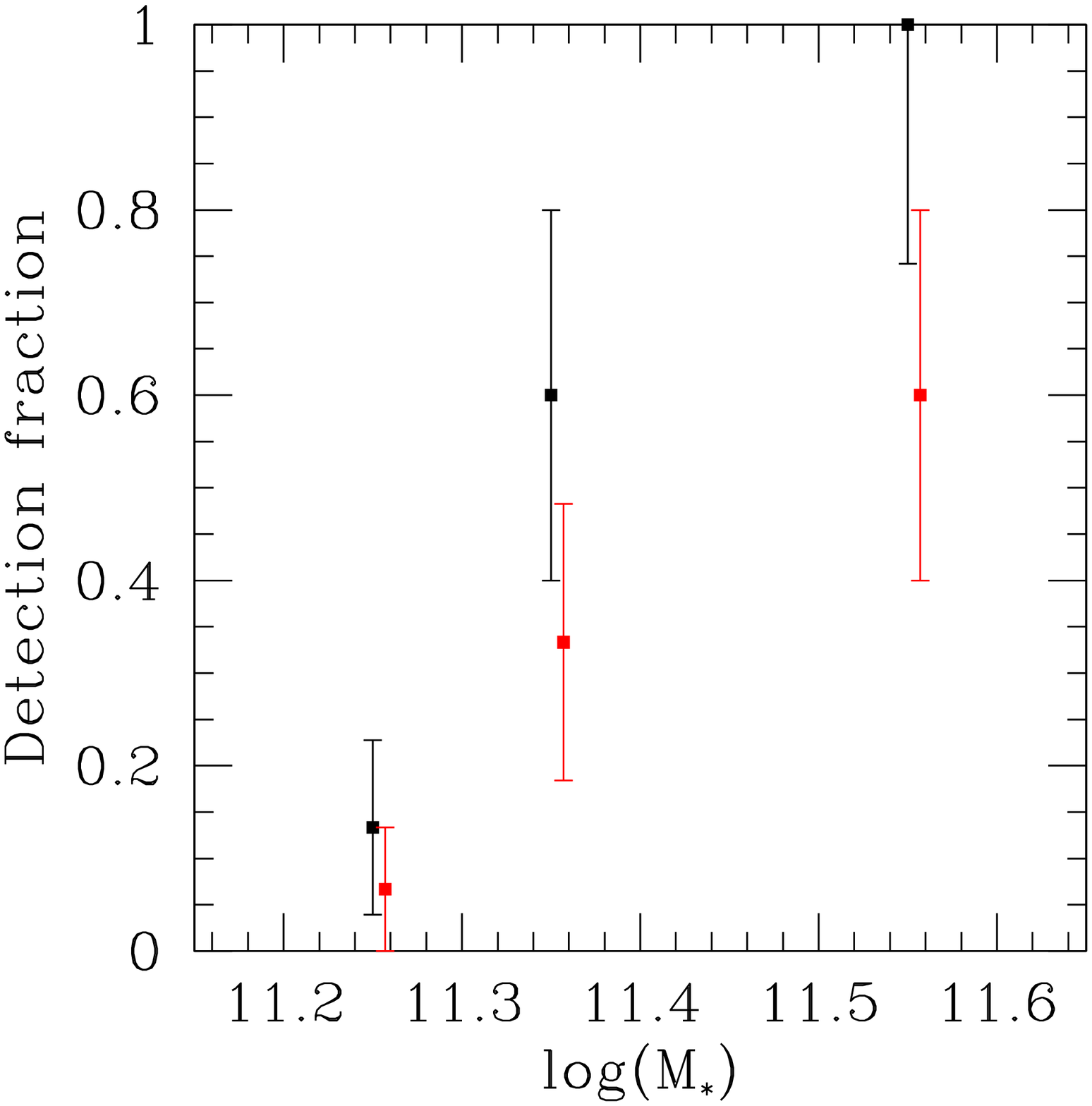}
\includegraphics[width=0.48\columnwidth]{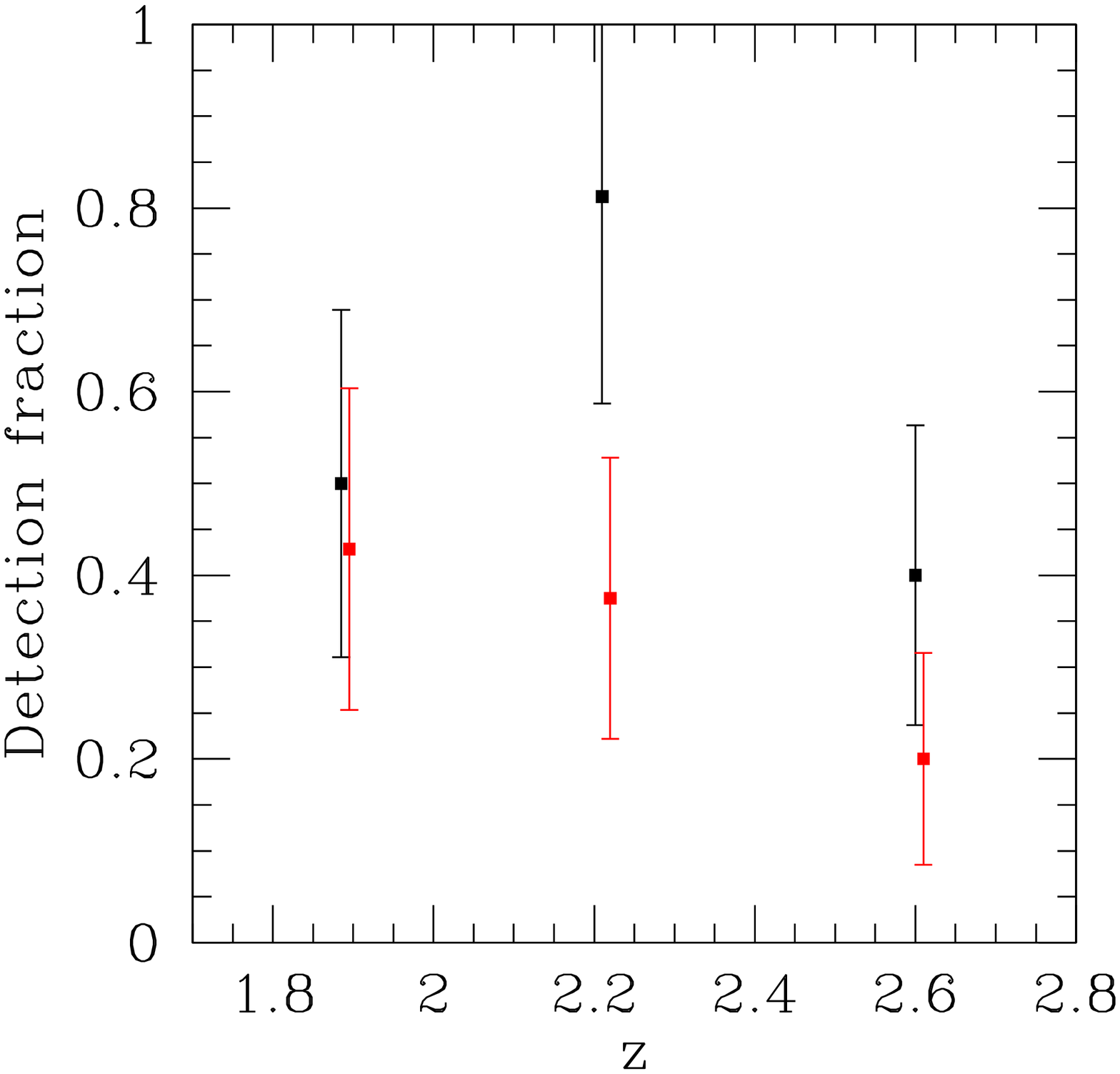}
\caption{Detection fraction as a function of S\'ersic index (upper left panel), effective radius (upper right, units of kpc), mass (bottom left) and redshift (bottom right).  The black dots indicate the 24~$\mu$m detection fraction with Poissonian error bars, while the red dots show the fraction of 250~$\mu$m detected sources.
}\label{DFnrm}
\end{figure}

In the four panels of Fig.\ref{DFnrm} we show the trends for the detection fractions for MIPS- and SPIRE-detected galaxies as a function of S\'ersic index (upper left panel), effective radius (upper right panel, units of kpc), stellar mass (bottom left panel, in Log(M$_{\odot}$)) and redshift (bottom right). The points represent the centres of  three equally populated bins in the parent sample. The black dots indicate the mean detection fractions with Poissonian error bars for MIPS, while the red dots are the fractions of SPIRE $250\mu$m detected sources. These panels indicate a trend in the sense of higher detection for large, massive, disk-like galaxies. 

In Table \ref{tab1} we summarise the results for the detection fraction, SFR and sSFR by dividing the sample according to morphology, for MIPS- and SPIRE-detected galaxies. The fractions of MIPS-detected galaxies are compatible with previous results (e.g. PG08). 

The mean SFR and sSFR values are derived by taking into account non-detections for SPIRE, as explained below. We assume as lower limit null SFRs for $250\mu$m non-detected galaxies. Upper limits are derived by assuming a  flux in the $250\mu$m band equal to the $5\sigma$ limit of 4.4mJy for all non-detected sources and performing a fit of the mean SED for the whole sample (and the disk-/spheroid-like galaxies sub-samples) of non-detections. The reported errors refer to the semi-interval between the upper and the lower limits.

When computing the MIPS SFR/sSFR values we take into account the systematic effects measured in recent {\it Herschel} works (see Elbaz et al. 2010; Nordon et al. 2010; Rodighiero et al. 2010), where it is found that MIPS based LIR estimates suffer of a systematic overestimation effect. The mean value for the ratio between 24$\mu$m and $250\mu$m based LIR measurements for our sample results to be $\sim3$, in agreement with the value found by Elbaz et al. (2010; see their Fig.~2) for high-z galaxies and using PACS+SPIRE data. Nordon et al. (2010) report a higher value (from 4 to 7 for high-z galaxies) but this could be related with the fact that in this latter work the factor is derived by fitting purely PACS data.

The mean values for the SFR of the sub-samples defined according to the morphological type indicate higher SFR for disk-like galaxies with respect to spheroidal-like galaxies. This behaviour is reflected in the mean sSFR values determined for the two sub-samples. Furthermore, the SFR estimates well match other literature results (e.g. PG08, Rodighiero et al. 2010, Viero et al. 2010), once we restrict to the same mass-redshift range and convert the results to the same IMF. 
In a forthcoming Paper we will perform a complementary full stacking analysis (and take advantage of the larger statistics provided by the full GNS sample) in order to derive more robust estimates.

\begin{figure}
\includegraphics[width=1\columnwidth]{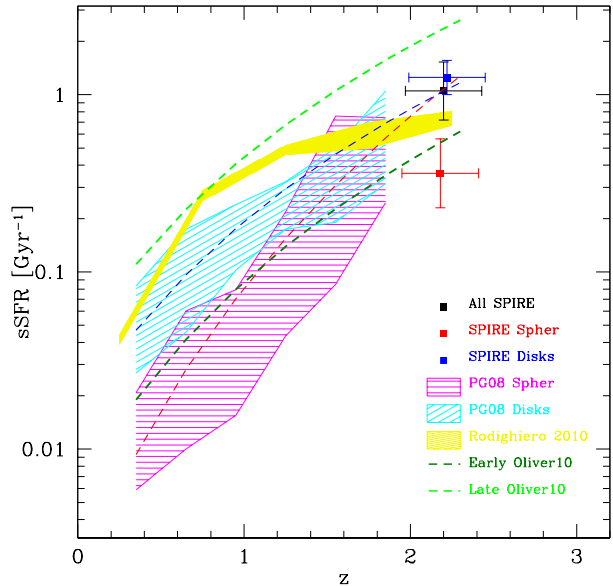}
\caption{Comparison of the mean sSFR at $z\simeq 2.3$ (corresponding to the median redshift of our sample) for SPIRE detected galaxies (black square for the global population, red and blue for disk- and spheroid-like galaxies, respectively) with other results on the evolution of the sSFR.  The magenta and cyan shaded regions show the sSFR measured by PG08 for early and late-type galaxies detected at $24\mu m$, respectively. The red dashed line and the blue dotted one are their fitting relations for the redshift evolution of the sSFR (see text for details). The light- and dark-green lines are the fitting relations for late- and early-type galaxies, respectively, in Oliver et al. (2010a).  The yellow region indicates the estimate of Rodighiero et al. (2010), using PACS data. 
}\label{ssfr}
\end{figure}

In Fig.\ref{ssfr}, we show the mean specific SFR of massive galaxies  detected by {\it Herschel}, as a function of redshift and morphology, compared to PG08, Oliver et al. (2010a) and Rodighiero et al. (2010). 
When segregating the sample based on the S\'ersic indices, PG08 find that the specific SFRs of massive ($M>10^{11}$M$_{\odot}$) spheroid galaxies detected at $24\mu$m evolve as  $(1+z)^{5.5 \pm 0.6}$ from $z = 0$ to $z = 2$ (red dashed line in Fig.\ref{ssfr}), while the evolution for disk-like galaxies detected at $24\mu m$ goes as $(1+z)^{3.6 \pm 0.3}$ (blue dotted line in Fig.\ref{ssfr}). 
Also showed are the mean relations for early- (dark green dashed line) and late-type (light green) galaxies obtained by Oliver et al. (2010a) analysing data from the Spitzer Wide-area InfraRed Extragalactic Legacy Survey (SWIRE; Lonsdale et al. 2003). 
The red dot indicates the mean sSFR for spheroid-like galaxies detected with SPIRE, while the blue dot is for the disks-like galaxies. The black point is the global mean value (the points are slightly shifted in $z$ for clarity). The yellow shaded region indicates the trend found by Rodighiero et al. (2010) using only {\it Herschel}-PACS (Photodetector Array Camera and Spectrometer, Poglitsch et al. 2010) data for a larger sample of GOODS-N galaxies (in the mass range $10^{11}$M$_{\odot}<M<10^{11.5}$M$_{\odot}$). 

\section{Discussion and conclusions}
In this paper we have analyzed the FIR properties of an optically selected sample of 45 high redshift ($1.7\le z \le 3$) massive ($M>10^{11}$M$_{\odot}$) galaxies as a function of their structural parameters.  Our analysis is based on the measurement of the specific SFR for each galaxy based on their IR emission, exploiting the deep {\it Herschel}-SPIRE data obtained by the HerMES Project in the GOODS North field.
Our main conclusions are the following:
\begin{itemize}
\item{ we globally detect $33$ per cent of the galaxies of the parent GNS sample at least in one SPIRE band, this is in agreement with the detection fraction found by Elbaz et al. (2010) analysing the whole GOODS-N data sample with PACS and/or SPIRE detections.  Most of the SPIRE detected sources in our sample are disk-like objects.  More precisely, the detection fraction of galaxies with $n \le 2$, is $45$ per cent,  while we detect only $\sim13$ per cent of the spheroid-like galaxies ($n > 2$).   Among the MIPS 24~$\mu$m detections, the detection fractions are in agreement with other literature results (e.g. PG08)}
\item {for both the MIPS- and SPIRE-detected fractions (see Fig.\ref{DFnrm}) a trend is present indicating that larger, massive, disk-like galaxies are better detected with respect to small, spheroidal, less massive galaxies. This would confirm the findings of Kriek et al. (2009) who find, using optical spectroscopy, that large disk-like galaxies are much more star-forming than compact spheroid-like galaxies}
\item{ as an additional proxy of the compactness of galaxies we have split our sample based on galaxy surface density. We find that on average less compact galaxies are more easily detected with respect to more compact objects. We also find a small indication of a larger mean size and mass for detected galaxies with respect to non-detections. This is supported by the trends shown in Fig.~2}
\item{ by means of fitting to the IR part of the spectral energy distribution, we have derived the SFR and sSFR for the SPIRE and MIPS sub-samples including the contribution from non-detections for SPIRE and accounting for the systematic effect of overestimation of the LIR when using MIPS24 based measurements. The two estimates result in good agreement} 
\item{ disk-like galaxies show a mean higher value ($\sim 3$ times) of SFR and sSFR with respect to spheroid-like galaxies for both SPIRE and MIPS based measurements.  Our estimated mean sSFR for the two morphological classes at $z \simeq 2.3$ well match the sSFR evolution of previous studies (PG08, Oliver et al. 2010a, Rodighiero et al. 2010).  
}
\end{itemize}
The debate on the nature of these different population of galaxies (compact and large galaxies) is still open. It is possible that they are linked through dissipative major mergers, as suggested by e.g. Ricciardelli et al. (2010), or they might follow two separate evolutionary paths and the differences in the structural parameters simply reflect different formation epochs.
In the first case, massive galaxies would go through a phase of high star formation and the transformation from isolated, gas-rich disk-like galaxies with typical sizes of $\sim 2-3$ kpc, into compact ($R_e \sim 1$ kpc) less star-forming galaxies is driven by violent major merger events, compatible with the scenario depicted by theoretical models. 
In the second case, massive galaxies at $z\ga 2$ must have formed very quickly, and consequently their high stellar densities could reflect the high gas densities in the primeval Universe. 
The high fraction of merging systems suggested by the inspection of ACS- and NICMOS-{\it HST} images for the objects presented in this work would go in the direction of supporting the first proposed scenario, but further investigations are necessary to confirm this hypothesis. 

This work represents the first step in the direction of understanding FIR properties of these high-$z$ compact massive galaxies and will be followed by a more extensive study exploiting the whole GNS sample and including the deep PACS observations for GOODS North and South fields provided by the PACS Evolutionary Project.   
\section*{Acknowledgments}
{\small 
Special thanks to E.~Ricciardelli, J.~Fritz and J.M.~Varela-Lopez for useful discussions and technical support.
The data presented in this paper will be released through the Herschel database in Marseille HeDaM ({\rm hedam.oamp.fr/herMES}).
SPIRE has been developed by a consortium of institutes led by
Cardiff Univ. (UK) and including Univ. Lethbridge (Canada);
NAOC (China); CEA, LAM (France); IFSI, Univ. Padua (Italy);
IAC (Spain); Stockholm Observatory (Sweden); Imperial College
London, RAL, UCL-MSSL, UKATC, Univ. Sussex (UK); Caltech, JPL,
NHSC, Univ. Colorado (USA). This development has been supported
by national fund- ing agencies: CSA (Canada); NAOC (China); CEA,
CNES, CNRS (France); ASI (Italy); MCINN (Spain); SNSB (Sweden);
STFC (UK); and NASA (USA). AC acknowledges a grant from the Spanish MCINN: ESP2007-65812-C02-02. 
}

\label{lastpage}
\bsp
\end{document}